# Discrimination between cosmological and stellar phenomena by the intensity interferometry

P. B. Lerner[1], N. M. Miskovsky[2], P. H. Cutler[3]

## Abstract

We provide a quantitative theory of discrimination between objects with the same color temperature but having different angular spectrum by intensity interferometry. The two-point correlation function of the black body image with extended angular spectrum has significant differences with a correlation function of a black body with a narrow angular spectrum.

1. **Introduction**

Submillimeter and terahertz astronomy is a quickly developing field with numerous other security, defense, and imaging applications (Walker, 2015), (Ricke, 2012), (H.-J. Song, 2015). In particular, Cosmic Microwave Background (CMB) radiation lies in the millimeter region of the electromagnetic spectrum and, because this spectrum is close to the black body, it extends far into the submillimeter region. However, submillimeter and terahertz astronomy suffers several deficiencies. First, these waves are efficiently absorbed by the atmosphere. This can be remedied by placing detection equipment in the high mountains, Antarctica, or the space probes—all expensive options. Second, both receivers and emitters—required for instance, for the local oscillators—for the terahertz are few. (Brundermann, 2011). The authors of the current paper proposed to use nanoantennas to accomplish both of these goals but, so far, progress has been insufficient. (P. B. Lerner, 2015) Finally because the blackbody radiation between a few and a hundred kelvin lies in the submillimeter/terahertz range, a spectral region that many space objects—planets, space dust, nebulae—have emissions. (Walker, 2015) So, the discrimination between different objects of interest presents a complicated problem. In particular, alleged observation of traces of inflation of the Universe in CMB spectra was later ascribed to the scattering from the space dust (BICEP2, 2018). Heretofore, the development of experimental methods, which distinguish between light from objects with a similar color temperature and polarization presents an important practical problem. An intuitive basis of our discrimination protocol is that the angular distribution of the CMB is a collective phenomenon, not localized in any particular region of the sky unlike that of stars and nebulae, which may have more localized distribution.

---
[1] Scholar-at-large, Anglo-American University, Prague and Device Consultants, LLC, pblerner@syr.edu, pbl2@psu.edu
[2] Professor (emeritus), Penn State University (Altoona) and Device Consultants, LLC
[3] Professor (emeritus), Penn State University and Device Consultants, LLC

The discrimination problem cannot be, in general, solved by heterodyning the signal with the local oscillator because the spectral characteristics of objects are nearly identical (Daher, 2016). Another problem is that what is considered "noise" in a recognition problem is not random and cannot be eliminated by averaging many snapshots of an object. As we shall see below, if anything, what we consider here as noise is usually more coherent than the image of the valuable signal. Finally, we use the scalar theory of light propagation, thus automatically assuming identical polarization of the light from the objects.

## 2. Van Cittert-Zernike Theory

The foundational quantity we have to estimate for the discrimination between objects of the same temperature, is a two-point correlation function. To demonstrate the method, it is sufficient to use scalar EM theory, though, the extensions of the Van Cittert-Zernike formula exist for the vector propagation (Ostrovsky, 2009), (Tervo, 2013). In J. W. Goodman, (Goodman, Speckle phenomena in Optics, 2007), Eq. (4-54), the Equation for the two-point correlation function for the monochromatic radiation takes the form:

$$\Gamma(x_1, y_1, x_2, y_2; k) = \frac{k^2}{16 \pi^2 z^2} \exp\left[\frac{ik}{2z}(x_1^2 + y_1^2 - x_2^2 - y_2^2)\right] \times$$
$$\iint I(X', Y') \exp\left[\frac{ik}{2z}\left(X'(x_1 - x_2) - Y'(y_1 - y_2)\right)\right] dX' dY' \quad (1)$$

Our simplified geometry of the problem is depicted in Fig. 1 (a more accurate picture can be found in Morse and Feschbach ( (Morse, 1953), §11.4, p. 1541)). In Equation (1), *I(X, Y)* is an intensity distribution in the object plane. Because $x_{1,2}$ and $y_{1,2}$ are the instrumental distances within the apparatus, we can neglect the difference between z—the transversal coordinate and r—the distance to the observed object. Then, Equation (1) can be rewritten in the form:

$$\Gamma(\rho, \theta, \varphi; k) = \frac{k^2}{16 \pi^2 z^2} \exp[ik(\cos\varphi \cos\theta \, \Delta x + \sin\varphi \cos\theta \, \Delta y)] \times$$
$$\iint f(\theta', \varphi') \exp[-ik(\cos\varphi' \cos\theta' \, \Delta x + \sin\varphi' \cos\theta' \, \Delta y] \, d\varphi' d\theta' \quad (2)$$

Or, more conveniently for computational purposes:

$$\Gamma(\rho, \theta, \varphi; k) = \frac{k^2}{16 \pi^2 z^2} \exp\left[\frac{ik\overrightarrow{r \cdot \rho}}{z}\right] \times$$
$$\iint f(\theta, \varphi') \exp\left[-\frac{ik\overrightarrow{r \cdot \rho'}}{z}\right] d\varphi_2 d\rho' \quad (2a)$$

In Equation (2), we introduced an angular spectrum of radiation by the formula:

$$I(X', Y') = \frac{I_0 f(\theta, \varphi)}{4\pi r^2} \approx \frac{C \cdot f(\theta, \varphi)}{4\pi z^2}$$

Where $C$ is the source intensity per *srad*. And used the notation: $\Delta x = x_1 - x_2$, $\Delta y = y_1 - y_2$. The radial coordinate $\rho$ is defined as

$\rho = \sqrt{\Delta x^2 + \Delta y^2}$. The parameter $\rho$ is the distance between two observation points in the focal plane, for which correlation function is measured. Further, refer to $\rho$ as "length".

Finally, because our object is non-monochromatic and incoherent, the final result of Equation (2) must be integrated over the spectral distribution of the incoming radiation (Akhmanov, 1981). Below we use the blackbody spectrum, which is represented by CMB with a great degree of accuracy (Mather, 1994).

Because we deal with the celestial sphere, it is conventional to expand angular spectrum into series over the spherical harmonics:

$$f(\theta, \varphi) = \sum_{n, |m| \le n}^{n, m = \infty} c_{nm} Y_{nm}(\varphi, \theta) = \sum_{n, m = 0}^{n, m = \infty} c_{nm} P_n^m(\cos\theta) e^{im\varphi}. \quad (3)$$

Then, the integrals over $\varphi_2$ and $\rho'$ in Equation (2a) can be easily computed:

$$\int_0^{2\pi} \exp[ik\rho' \cos(\varphi - \varphi_2) + im\varphi_2] d\varphi_2 = i^m J_m(k\rho')$$

And

$$\int_0^{\infty} J_m(k\rho') d\rho' = \frac{1}{k}$$

The resulting equation for the k-component of the transversal correlation function is then expressed as follows:

$$\Gamma_{\perp}(\rho, \theta, \varphi, \varphi_1; k) \propto \frac{k}{16\pi^2} \sum_{n,m} (-1)^n c_{nm} P_n^m(\cos\theta) \times \\ \exp[-ik\rho(\cos(\varphi - \varphi_1) + im(\varphi - \varphi_1)] \quad (4)$$

Where we omitted the average intensity of the source. Using the blackbody density of states, we obtain for the transversal correlation function the following Equation:

$$B_\perp(\rho,\theta,\varphi) = 2 \iint u_k\, \Gamma_\perp(\theta,\varphi,\varphi_1;k)k^2 dk d\varphi_1 = \frac{1}{8\pi^2}\Sigma_{n,m}(-1)^n c_{nm} P_n^m(\cos\theta)\exp(im\varphi)$$
$$\iint \frac{k^3 \cdot \exp(-ik\rho\cos\varphi_1 - im\varphi_1)}{\exp(\alpha k)-1} dk d\varphi_1 \tag{5}$$

The angular integral is again easy to compute, which produces the formula:

$$B_\perp(\rho,\theta,\varphi) = \frac{1}{8\pi^2 \alpha^4}\Sigma_{n,|m|\le n}(-1)^n c_{nm} P_n^m(\cos\theta)\exp(im\varphi) \times$$
$$\int \frac{x^3 J_m(x\frac{\rho}{\alpha})}{\exp(x)-1} dx \tag{5a}$$

Equation (5a) is our main equation for computing the transversal correlation function of the arguments θ and φ—the direction of the source and ρ—the base of the interferometer.

### 3. "Toy" Problem I

An intuitive reason for the differences in the two-point correlation function of extended and compact objects with the same color temperature is that the Van Cittert-Zernike theorem has as its expression a modified Fourier transform. Henceforth, an object consisting of relatively narrow angular features has a lot of spatial frequencies, while an object with an extended angular spectrum has different spatial harmonics averaging out and a relatively compact range of spatial frequencies. This particular fact can be verified from Equation (5) in the limiting cases. For the small values of an integration angle $\varphi_1$, the phase under an integral sign oscillates fast with a growing value of $k \cdot \rho$. Otherwise, the value of cosine is small, and the correlation function extends for a longer distance.

Herewith, we demonstrate the difference in the two-point correlation functions of the extended (Fig. 2) and compact (Fig. 3) "objects", by which we represent "cosmological" and "stellar" or "nebular" phenomena, respectively. Naturally, these objects are just the digitized photographs obtained from the nasa.org site. In Fig. 4 and Fig. 5, we observe the visual difference between the plots of two-point correlation functions computed by Equation (5a).

To convert an artificially colored image into a digital file we use the following preprocessing procedure. Namely, we digitize each of the three basic colors and use a linear combination of each pixel with coefficients, which square to one, and which has a maximum variance. We assume that this superposition of basic colors provides the most informative pixilation at least among the linear combinations.

Our digitization procedure assumes that Fig. 2 and other images are a Mercator projection of the solid angle, i.e. the square map subtends the polar angle within [0, 2π] and the azimuthal angle from [0, π]. This is, of course, a crude approximation but more accurate rasterizing is not necessary to demonstrate the essence of the method.

The main numerical problem with the Equations (5-5a) is that, despite their ubiquity in theoretical demonstrations, a series of spherical functions display remarkably slow convergence in the case of irregular patterns. Henceforth, one needs relatively uneconomical sets of spherical functions of Equation (3) to faithfully reproduce even a modest numerical image (500k-1M pixels).[4] To make the situation manageable for a PC, we use *n=12* and *m=[-12,12]* in Equation (5a). To gauge the angular resolution achieved with this approximation, we use a naïve formula, which is a consequence of Shannon-Nykvist-Kotelnikov[5] theorem for the angular harmonics:

$$\Delta\phi = \frac{180°}{n} \approx 15°$$

This resolution is comparable to the angular dimension of the overlapping star cluster (Fig. 6).

We notice that the magnitude of short-length ($\rho/\lambda_T \leq 10$) two-point correlations is 1-2 orders of magnitude higher for a compact object than for an extended object, which very much complicates a signal-to-noise resolution.

## 4. "Toy" Problem II

In this section, we demonstrate the correlation functions of synthetic objects created by an overlap of a stellar cluster and CMB background. One synthetic object was created by superimposing a cluster image on the CMB temperature map (Fig. 6), another—by numerically adding up 80% of the contrast of the CMB map and 20% of the star cluster. We distinguish these cases as IIa and IIb.

We demonstrate the results of the simulation in the first case in Fig. 7. Compared to Fig. 4, the scale of the oscillations of the two-point correlation function is larger by an order of magnitude and its behavior as a function of azimuthal angle is drastically different for the short differences in the spatial argument ρ. The result of the numerical overlap of pictures is similar, yet it is less transparent for interpretation because delta-like spikes—the result of rapidly changing contrast—are ignored by its numerical approximation (Fig. 8).

Heretofore, we notice that the behavior of the two-point correlation function can be quite different depending on whether the object has an extended angular spectrum, which we identify with cosmological phenomena or a sharply changing contrast typical of nebular or stellar objects.

---

[4] This volume of data is very small by modern standards of modern digital processing, yet, it pretty well corresponds to the resolution of the modern astrophysical cameras. (BICEP2, 2018).

[5] Shannon-Nykvist-Kotelnikov theorem (Goodman, Introduction to Fourier Optics, 2005)

## 5. Filtering problem

Once we observed that there is a physical characteristic—two-point and, probably, higher correlation functions—which are different for the signal and noise of the same color temperature, we can try to discriminate between them. The problem of filtering out a stellar image (noise) from the background radiation (signal) does not only concern signal-to-noise ratio (SNR) which is typically below 1. Another complication is that statistical characteristics of the signal and noise, in this case, are the opposite to normal—noise has a relatively long coherence length, while the coherence length of a signal is relatively small. If the correlation function of the noise were flat, there would be a little difficulty to filter it out but it oscillates with amplitude higher than the amplitude of the signal.

One proposed method of filtering is based on the cutoff of higher angular harmonics in the combined image. This method does not use any previous knowledge about the "useful" signal. Our example deals with attenuation of the higher harmonics by a Gaussian factor in the Equation (5a):

$$\tilde{B}_\perp(\rho,\theta,\varphi) = \frac{1}{8\pi^2\alpha^4}\sum_{|m|\leq 12}(-1)^m c_{12,m} P_{12}^m(\cos\theta)\exp(im\varphi) \times \exp(-\rho^2 m/2\sigma^2)\int \frac{x^3 J_m(x\frac{\rho}{\alpha})}{exp(x)-1}dx \qquad (6)$$

Where σ is the smoothing parameter of the dimension length. By varying this parameter, one can suppress higher angular harmonics at shorter distances where they seem to mask a useful signal. Generally, the expression $B_\perp(\rho,\theta,\varphi) - \tilde{B}_\perp(\rho,\theta,\varphi)$ can behave itself quite irregularly but at short distances, most of the contribution comes from an original object.

The above method being taken literally produces uncertain results. Only in one case of Toy Problem IIb—a simple linear combination (0.8:0.2) of digitized images of the CMB and nebula—a filtered image has a significantly higher positive correlation with the image of Fig. 5 (see Table 1). In that optimum case (σ=1.5α), the correlation between the filtered correlation function and the correlation function of the true image grows approximately 2.5 times with respect to the baseline case.

A variation on this method using some a priori information would be to approximately match the asymptotic behavior of the correlation function in coordinate ρ of the filtered image with the expected asymptotic for the CMB without the star. The results are shown in Table 2. To estimate how significant the difference in correlation between the unfiltered and filtered images is, we assume that the correlations in Table 2 are the sample averages. Then, using Fisher z-transformation and the same hypothetic size of the unfiltered and filtered sample, we can compute z-score for the two-point correlation function as a function of an azimuthal angle φ ( (Kanji, 1999), Test 14). Then, for the n=1000 images, the score will be z=-2.33 (probability of two-tail null, P=0.02) and for n=6000 images, the score will grow to z=-7.39 (P<$10^{-6}$).

**Table 1**. Coefficient of correlation of 400-pixel two-point correlation function of the CMB object with the same object overlapped by a star image.

| Case | Toy IIa, $\tilde{B}_\perp(\rho,\theta,0)$. | Toy IIa, $\tilde{B}_\perp(\rho,0.11\pi,\varphi)$. | Toy IIb, $\tilde{B}_\perp(\rho,\theta,0)$ | Toy IIb, $\tilde{B}_\perp(\rho,0.11\pi,\varphi)$ |
|---|---|---|---|---|
| No filtering | 0.789 | -0.03 | -0.514 | 0.128 |
| Smoothing parameter, α | | | | |
| 0.5 | 0.381 | -0.619 | -0.395 | 0.222 |
| 1.5 | 0.418 | -0.391 | -0.387 | **0.299** |
| 5. | 0.726 | -0.115 | -0.454 | 0.159 |
| 15. | 0.787 | -0.044 | -0.507 | 0.132 |
| 50. | 0.789 | -0.035 | -0.513 | 0.128 |

**Table 2**. Correlation of 400-pixel two-point correlation object with the fitted long-range asymptotic behavior (σ=15α) in the Toy IIa case.

| Correlation with the original image | Unfiltered | Filtered |
|---|---|---|
| $\tilde{B}_\perp(\rho,\theta,0)$. | 0.813 | 0.809 |
| $\tilde{B}_\perp(\rho,0.11\pi,\varphi)$. | 0.521 | **0.593** |

**Conclusion**

The behavior of the two-point correlation function in the cases of compact and extended objects, (or smooth and ragged objects) with the same color temperature for their discrimination is quite different. Since 1960s (Bourret, 1960), (Kano, 1962), (Mehta, 1967), it has been known that blackbody radiation is not delta-correlated but has a pronounced coherence at the scale $r_{corr} \sim \frac{\bar{\lambda}}{\Delta\theta}$, where $\bar{\lambda} = \frac{2\pi\hbar c}{k_B T}$ is a characteristic wavelength of the blackbody radiation and $\Delta\theta$ is the characteristic width of the angular spectrum of an object. This information can be used to distinguish between compact and extended objects.

The direct measurement of the second-order coherence for the simplest possible object—solar radiation—was accomplished only in 2012 (Mashaal H., 2012). We propose to use two-point and, possibly, higher correlation functions for the discrimination of the cosmological phenomena and nebular phenomena. The distinction is that the effective radiator in the cosmological case extends over complete solid angle in the observable sky, while in the stellar/nebular case, the radiator has a limited angular spread.

The main complication of the method is that a numerical value of the correlation functions of stars and nebulae seems to be larger by an order of magnitude than the correlation of the CMB for all, but a few directions. Yet, our estimates demonstrate that it would be possible to distinguish between overlapping structures extended for the whole sky and the images within a relatively small solid angle (5°-15°) through a sensible number (several thousand) of repeated measurements.

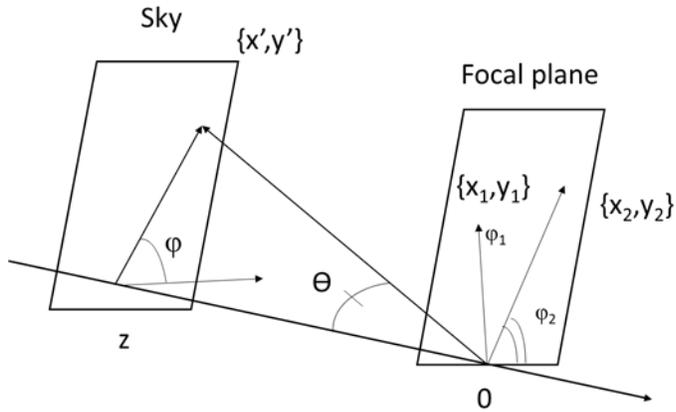

Fig. 1. The system of coordinates used in Section 3.

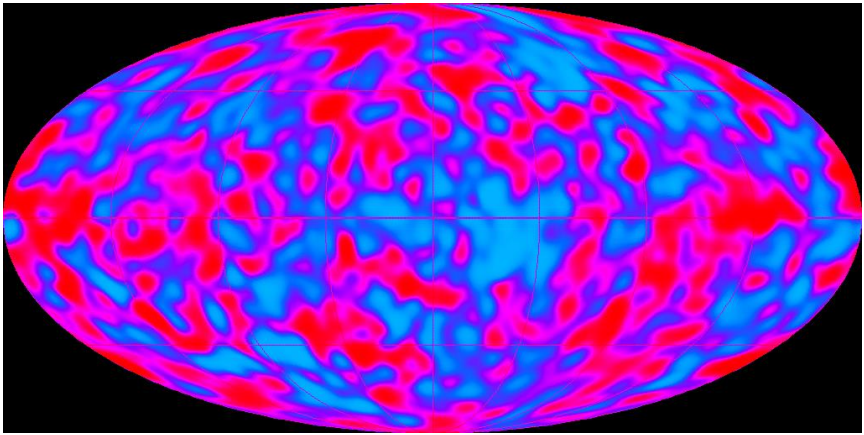

Fig.2. Temperature map of the Cosmic Microwave Background (CMB) by the COBE mission. The source: https://lambda.gsfc.nasa.gov/product/cobe/.

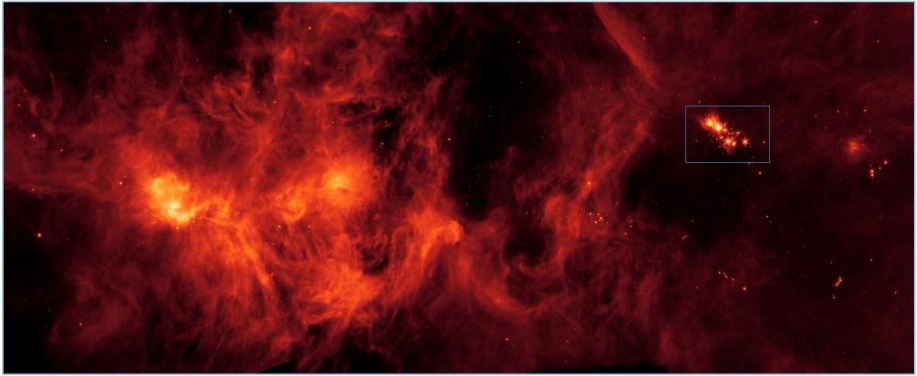

A)

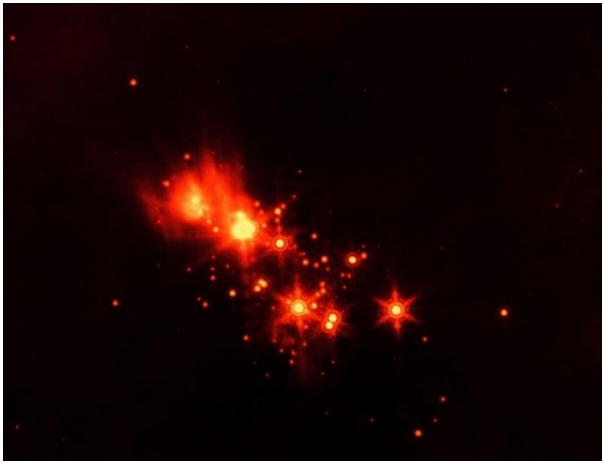

B)

Fig. 3. Perseus nebula. A) General view. A faint blue square indicates a star cluster used as our example of the stellar object. B) Perseus's inner cluster is used as a model of a compact object of stellar origin. The source of an original image: https://www.jpl.nasa.gov/images/perseus-molecular-cloud.

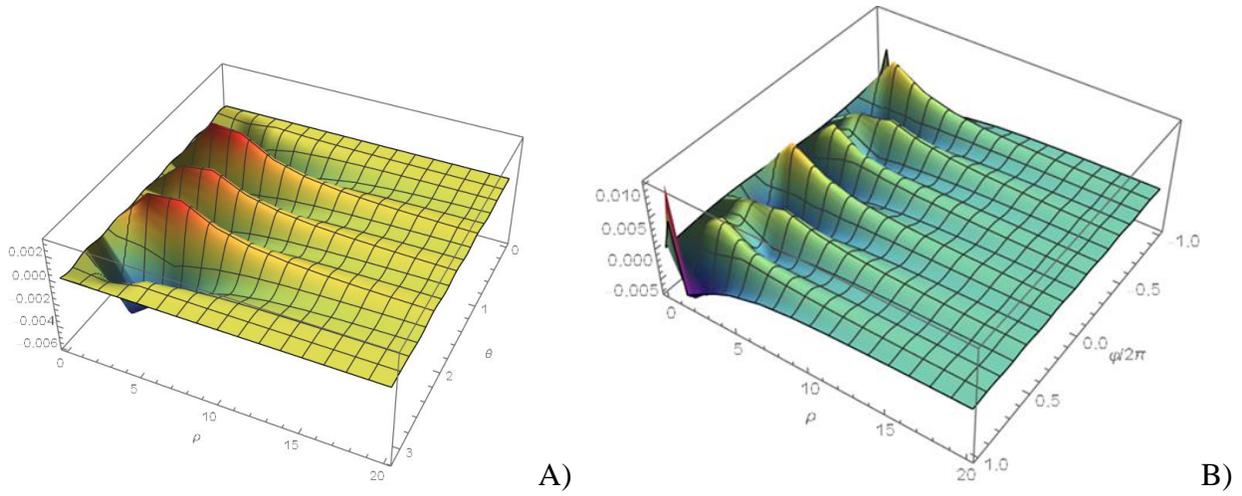

Fig. 4. Two-point correlation function B⊥(ρ, θ, φ ) of the CMB image in Fig. 1 according to Equation (5a). A) B⊥(ρ, θ, φ=0), B) B⊥(ρ, θ=0.5, φ).

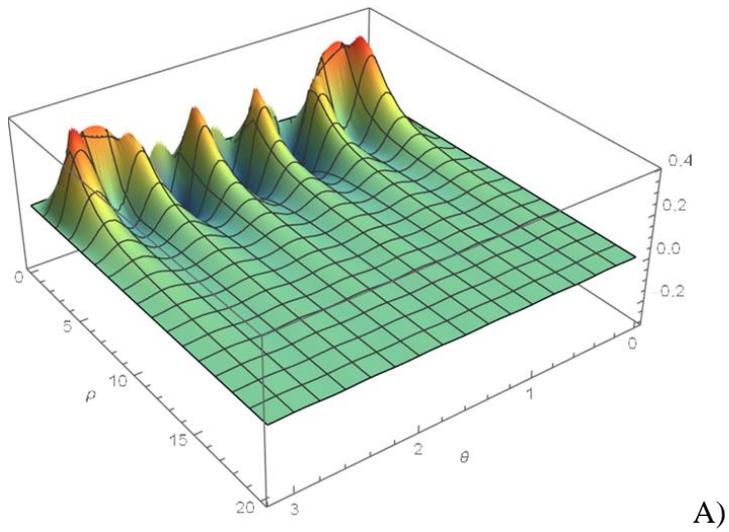

A)

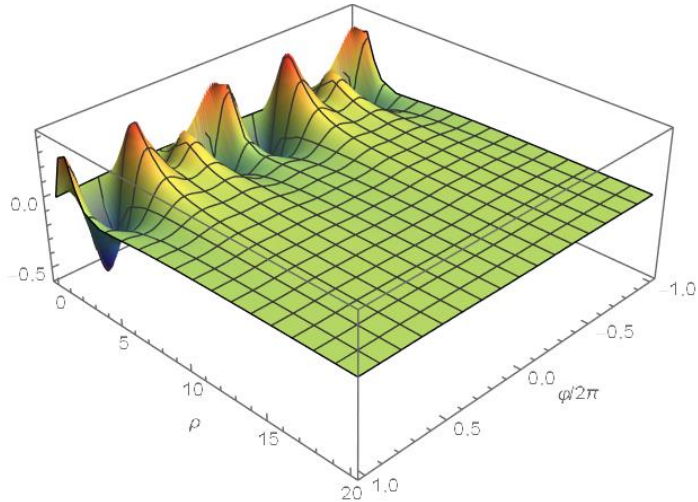

B)

Fig. 5. Two-point correlation functions of the compact object of Fig. 3 in the conditions of Fig. 4. The vertical scale of the correlations is 1-2 magnitudes higher and, consequently, correlations are observable at a longer range. The weakness of far-out oscillations compared to Fig. 4 is a visual artifact created by the change of scale of the drawings by approximately two orders of magnitude.

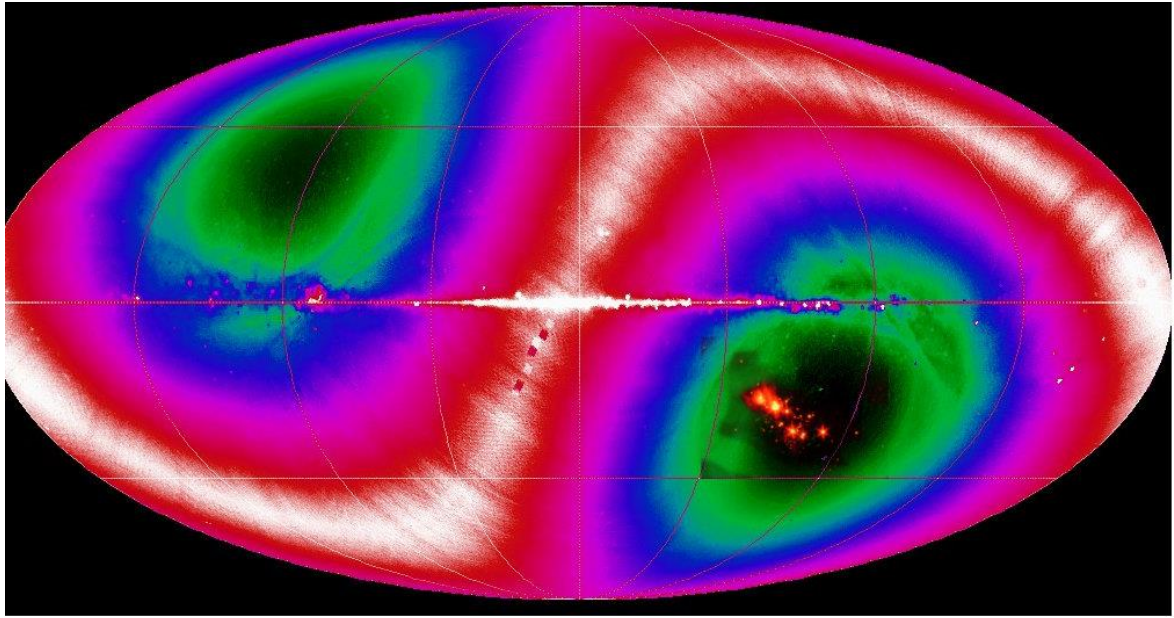

Fig. 6. The temperature map of the cosmic background radiation at 25 micron from the site https://lambda.gsfc.nasa.gov/product/cobe/slide_captions.cfm overlapped by an image of the star cluster from Fig. 2 in the lower left quadrant.

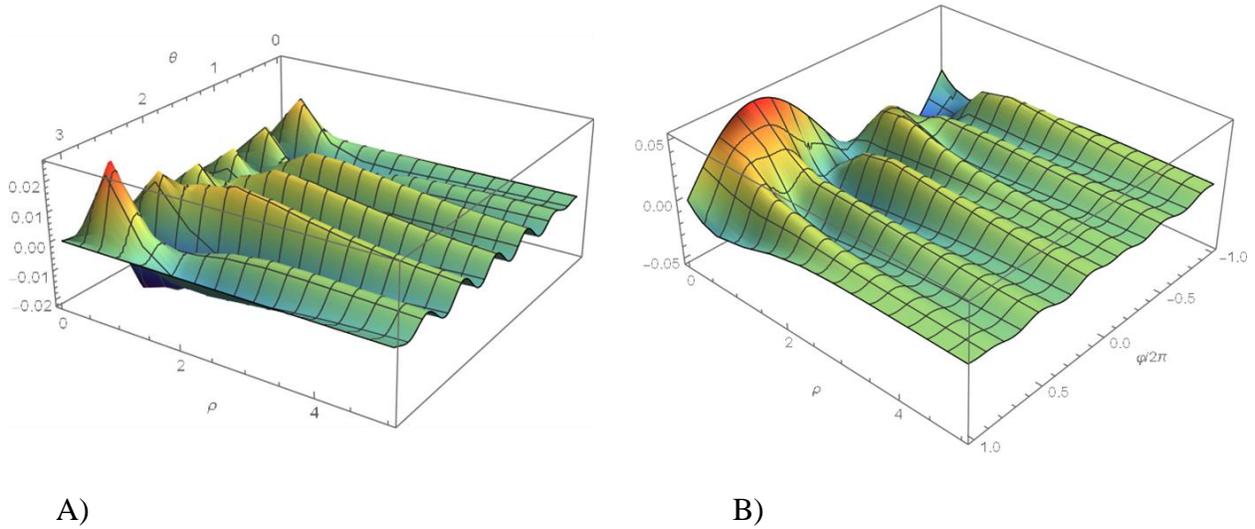

A)                                                         B)

Fig. 7. Two-point correlation functions of a composite object in Fig. 6 (for definitions see Fig. 3). Compared to Fig. 4B, Figure 6B displays a larger magnitude scale and different short-distance behavior. A large short-range feature in Figure 7B, is most likely the consequence of the central lobe in Fig 6 (the plane of the Milky Way).

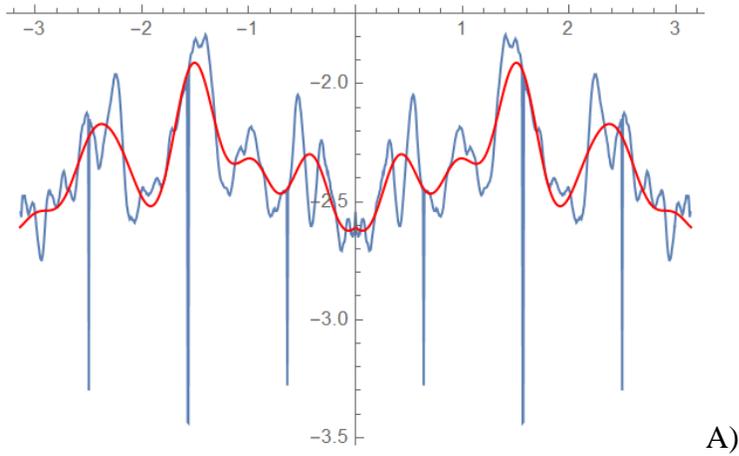

A)

Fig. 8. An example of discontinuous integrated map of the image for the azimuthal angle [-2π, 2π] in the Toy II case.